\documentclass[preprint,1p,11pt]{elsarticle}
\usepackage{mathrsfs}
\usepackage{epsfig}
\usepackage{amsmath}
\usepackage{amssymb}
\usepackage{slashed}
\newcommand{\bea}{\begin{eqnarray}}
\newcommand{\eea}{\end{eqnarray}}
\newcommand{\bnn}{\begin{eqnarray*}}
\newcommand{\enn}{\end{eqnarray*}}
\newcommand{\be}{\begin{equation}}
\newcommand{\ee}{\end{equation}}

\journal{Journal of Mathematical Analysis and Applications}
\def\MSC{\par\leavevmode\hbox {\it MSC:\ }}%

\begin{document}
\begin{frontmatter}

\title{Lie sphere geometry in nuclear scattering processes}

\author{S. Ulrych}
\address{Wehrenbachhalde 35, CH-8053 Z\"urich, Switzerland}

\begin{abstract}
The Lie sphere geometry is a natural extension of the M\"obius geometry, where the latter is very important in string theory and the AdS/CFT correspondence.
The extension to Lie sphere geometry is applied in the following to a sequence of M\"obius geometries, which has been investigated recently in a bicomplex matrix representation.
When higher dimensional space-time geometries are invoked by inverse projections starting from an originating point geometry, the Lie sphere scheme provides a more natural structure of the involved Clifford algebras compared to the previous representation.
The spin structures resulting from the generated Clifford algebras can potentially be used for the geometrization of internal particle symmetries.
A simple model, which includes the electromagnetic spin, the weak isospin, and the hadronic isospin, is suggested for further verification.
\end{abstract}



\end{frontmatter}

\section{Introduction}
The spin of a particle can be understood in terms of irreducible representations of the Poincar\'e group.
Thus, it has a clear geometric background. Since the success of the concept of isospin in the middle of the last century
physicists and mathematicians consider a possible geometrization of the internal symmetries of a quantum particle, see for example Hermann \cite{Her66}.
Today there is the hope that the geometrization problem can be addressed by string theory and the AdS/CFT correspondence \cite{Mal98,Gub98,Wit98}.
The acronym CFT indicates that the conformal symmetry plays a substantial role in these considerations.
However, in principle other geometries and symmetries could be considered. Even if one takes a step back into
a simplified world which consists only of two space dimensions, there are overall 23 families of Klein geometries \cite{Kom93}.

Among these geometries one finds the Lie sphere geometry \cite{Lie72}, which has been considered in detail for example by Cecil and Chern \cite{Cec87, Cec08}, Cecil and Ryan \cite{Cec15},
Sharpe \cite{Sha97}, Benz \cite{Ben12}, and Jensen et al. \cite{Jen16}. There is also the software library 
of Kisil \cite{Kis18} for graphical representation.
The Lie sphere geometry has been applied in the last decades in the study of Dupin submanifolds \cite{Dup22}. This research has been initiated by the thesis of Pinkall \cite{Pin81}.
The Lie sphere geometry considers points and lines as special cases of spheres, which can again be understood
as points in an ambient space.
Lie spheres have been applied in physics by Bateman, Timerding, and Cunningham within optics and electrodynamics \cite{Bat09, Bat10, Cun10, Tim12}. The radius of the Lie sphere is identified in these applications with $r=ct$.
This connection between geometry and relativistic physics has been investigated also by Cartan \cite{Car12}.
Thus, it seems to be worth to consider the Lie spheres for further applications. 
The following discussion therefore suggests to generalize the sequence of M\"obius geometries introduced in \cite{Ulr17} and apply the Lie sphere geometry instead of the M\"obius geometry.
From the perspective of physics one can then address the question mentioned in the beginning of this introduction
and try to identify the isospin symmetries of electroweak and strong interactions within the algebraic spin representations of a hierarchy of Lie sphere geometries.

The importance of projective geometry in relativistic physics has been noted already by Klein \cite{Kle10}.
Today string theory and the AdS/CFT holography point towards the importance of projective geometry in the representation of the fundamental geometries in physics.
Furthermore, one may assume that the dimensionality of physical space is not given a priori and the emergence of space-time has to be considered in a generalized geometric context \cite{Sei07,Zaa15}.
In AdS/CFT and string theory the geometry is generalized to higher dimensional spaces by adding space dimensions.
A constraint on the total number of dimensions has been provided by supergravity \cite{Nah78,Duf98}.
In \cite{Ulr17} and in the following discussion the next level of dimensionality is reached by adding one space and one time dimension, which is in line with the conformal compactification.
This guarantees that the concepts of projective geometry can be applied without modification within the series of geometric spaces of different dimension.
However, the special importance of Minkowski space-time cannot be derived in a natural way at this level of investigation.

Compared to its predecessor \cite{Ulr17} the following discussion, which is based on the Lie sphere geometry instead of the M\"obius geometry, provides a few advantages.
As shown in Section \ref{maks} the generalized approach is more natural with respect to the signature of the applied Clifford algebras.
The number of basis elements with positive square is equal to the number of basis elements with negative square.
In addition, the geometry originates by inverse projections out of a line instead of a plane, which is discussed in Section \ref{origin}. The line can be identified in its projective context with a point.
One may think here of the point geometry as an origin, which is inherent to the physical space with an a priori
undefined dimensionality. 
Section \ref{spinmomop} outlines how the content of \cite{Ulr17} is included as a subalgebra in the generalized representation.
In Section \ref{Lie} the algebraic representation is connected with the Lie sphere geometry.
The mass formula for electrons and protons, which has been discussed already in \cite{Ulr17},
indicates the emergence of particle configurations from their geometric context.
Finally, Section \ref{nuclear} shows that the approach is, compared to \cite{Ulr17}, more consistent with nuclear physics,
because the electromagnetic spin, the weak isospin, and the hadronic isospin can be assigned to the Clifford matrix representation of the proposed geometry.
The approach therefore appears to be more suitable than \cite{Ulr17} for the geometrization of internal particle symmetries.

\section{Bicomplex numbers}
The bicomplex numbers have been discussed by Segre \cite{Seg92} in 1892.
An equivalent number system called tessarines has been introduced by Cockle \cite{Coc48} even in 1848.
Detailed introductions with more information on the history of bicomplex numbers
can be found for example in \cite{Roc04,Lun15}. The bicomplex number can be further extended to multicomplex numbers, see Price \cite{Pri91}. An application
of multicomplex numbers in physics has been proposed recently in \cite{Cen19}. The bicomplex number
has one real and three imaginary parts
\be
\varsigma=\varsigma_1+i\varsigma_2+j\varsigma_3+ij\varsigma_4\,.
\ee
Both complex units of the bicomplex numbers are defined to square to minus one
\be
i^2=j^2=-1\,,\quad ij=ji=\sqrt{1}\,.
\ee
The two commutative bicomplex units are distinguished by their behavior with respect to the Clifford conjugation, which can be denoted in a physical context as spin conjugation. 
The complex unit $i$ changes sign, the bicomplex unit $j$ not
\be
\label{conjugation}
\bar{\varsigma}=\varsigma_1-i\varsigma_2+j\varsigma_3-ij\varsigma_4\,.
\ee
There is a second Clifford involution denoted as reversion. 
In physical applications one can identify this involution with Hermitian conjugation.
The reversion applied to the bicomplex number is changing the sign of both complex units
\be
\label{reversion}
\varsigma^\dagger=\varsigma_1-i\varsigma_2-j\varsigma_3+ij\varsigma_4\,.
\ee
The notation is inspired by the conjugation of Dirac spinors $\bar{\psi}=\psi^\dagger \gamma_0$.
Two complex units with different behavior with respect to conjugation can be applied potentially in the description of the CP violation,
where weak and strong phases have to be represented with exactly this property \cite{Gri08,Ger17,Lan17}.

The sign change of the hypercomplex units with respect to conjugation and reversion is consistent with previous publications of the author, which were based on hyperbolic numbers, see for example \cite{Ulr05}.
The hyperbolic complex numbers are defined slightly different, but they are in fact congruent.
The hyperbolic unit corresponds in terms of the bicomplex numbers to
\be
j_{HC}\equiv ij_{BC}\,.
\ee
$HC$ refers here to hyperbolic complex numbers and $BC$ to bicomplex numbers.
In the context of the following discussion it is worth to note that
the Lie sphere geometry in combination with hyperbolic numbers has been considered by Bobenko and Schief \cite{Bob16}.

The bicomplex numbers stand in relation to the sphere $S^3$ as the complex numbers are in relation to $S^1$.
In the following the radius of the bicomplex number will be set to $r=1$. This shortens the notation and takes into account that
in the projective context the length is an irrelevant quantity. The radius can be reintroduced if necessary.
Due to the diffeomorphism between $S^3$ and the group $SU(2,\mathbb{C})$ it is possible to represent
the bicomplex number in terms of the most general form of a group element of 
$SU(2,\mathbb{C})$
\be
\varsigma=e^{i\varphi}\cos{\vartheta}+je^{i\eta}\sin{\vartheta}\,.
\ee
In order to further study the properties of the bicomplex numbers one can also choose
the spin representation of $SO(3,\mathbb{R})$
to parametrize the bicomplex number
\be
\label{parmspin}
\varsigma(\alpha\beta\gamma)=e^{-\frac{i}{2}(\beta+\gamma)}\cos{\alpha/2}+je^{\frac{i}{2}(\beta-\gamma)}\sin{\alpha/2}\,.
\ee
One can now calculate the impact of conjugation and reversion on this representation.
Especially interesting is the squared modulus, which is usually leading to the squared radius of a complex number under the standard complex conjugation. 
For conjugation of the bicomplex number given by Eq.~(\ref{parmspin}) one finds with the help of Eq.~(\ref{conjugation})
\be
\label{consquare}
\varsigma\bar{\varsigma}=\cos{\alpha}+j\sin{\alpha}\cos{\beta}\,.
\ee
The map $\varsigma\bar{\varsigma}$ does not provide the squared radius and is even not a real number. 
A corresponding calculation can be done with the reversion. One finds by means of Eq.~(\ref{reversion})
\be
\label{revsquare}
\varsigma\varsigma^\dagger=1+ij\sin{\alpha}\sin{\beta}\,.
\ee
Here the squared radius appears as the real part, which corresponds in this case to $r^2=1$.
The remaining three terms in Eqs.~(\ref{consquare}) and (\ref{revsquare}) can be understood as the result of the
Hopf map $S^3\rightarrow S^2$. They form a parametrization of the sphere $S^2$. The expressions do not depend anymore on $\gamma$, which refers to the $S^1$ fiber of the Hopf map.

The relation between Pauli spin and the bicomplex numbers has been discussed before for example by Smirnov \cite{Smi07},
see also the references in this article for some applications in physics.
In this sense the representation of the bicomplex numbers as given by Eq.~(\ref{parmspin}) can be used to define the Pauli spinor as
\be
\chi_\frac{1}{2}=\varsigma(\theta\varphi 0)\,.
\ee
As discussed by Smirnov \cite{Smi07} the non-commutative Pauli matrices can be replaced by a new type of operators acting on the commutative bicomplex numbers.
As a consequence, the Pauli spinor with negative magnetic quantum number can be deduced from the first Pauli spinor by means of Hermitian conjugation and multiplication by the unit $j$
\be
\chi_{-\frac{1}{2}}=j\chi_\frac{1}{2}^\dagger\,.
\ee
One could think here of resolving all common spin structures, like the Dirac spin
or the isospin, into commutative number systems. If necessary, further multicomplex units can be introduced.
The following discussion will not follow this route, but uses a mixture of bicomplex numbers and conventional matrix representations, which are displayed in terms of non-commutative units,
in order to be aligned with \cite{Ulr17}. 

In the terminology of Clifford algebras the bicomplex numbers can be understood as the complexification $\mathbb{C}_{1,0}$
of the hyperbolic numbers $\mathbb{R}_{1,0}$. The basis of the real algebra is given by $e_\mu=(1,e_1)=(1,ij)$. 
The algebra is complexified by means of either $i$ or $j$.
Alternatively, the bicomplex numbers can be interpreted as the complexification $\mathbb{C}_{0,1}$ of the complex numbers $\mathbb{R}_{0,1}$.
The basis of the real algebra can be chosen as $e_\mu=(1,e_1)=(1,i)$. The complexification is then done with the unit $j$.

\section{Hypercomplex units for the representation of the Lie algebra $\mathfrak{sl}(2,\mathbb{R})$}
In \cite{Ulr17} it turned out to be useful to represent the Lie algebra $\mathfrak{sl}(2,\mathbb{R})$ with non-commutative hypercomplex units.
This provides additional insights into the structure of higher dimensional Clifford algebras 
and their geometries. The matrices are introduced as
\be
\imath=\left(\begin{array}{cc}
0&1\\
-1&0
\end{array}\right),\quad \jmath=\left(\begin{array}{cc}
0&1\\
1&0
\end{array}\right).
\ee
The two generating matrices are denoted by $\imath$ and $\jmath$.
Multiplication of the two elements results in
\be
\imath\jmath=\left(\begin{array}{cc}
1&0\\
0&-1
\end{array}\right)=-\jmath\hspace{0.03cm}\imath\,.
\ee
The three matrices correspond to the Lie algebra of the special linear group $SL(2,\mathbb{R})$. They will be used as the building blocks
for the sequence of Clifford algebras, which will be applied in the following sections.

\section{Hypercomplex representation of the Maks periodicity of Clifford algebras}
\label{maks}
The Clifford algebra periodicity, which has been investigated from a mathematical point of view by Maks \cite{Mak89}, will be applied in a hypercomplex representation similar to \cite{Ulr17} in order to generate higher dimensional geometries by inverse projections.
However, compared to \cite{Ulr17} a different set of Clifford algebras is involved.
The base geometry has now an odd number of $2m+1=n$ dimensions.
The $n-1$ basis elements of the paravector Clifford algebra $\mathbb{R}_{m,m}$ are transformed to the basis elements of the ambient Clifford algebra by
\be
\label{basspace}
e_{k}=\imath\jmath \otimes e_k\,,\quad k=1,\dots,n-1\,.
\ee
On the right-hand side of the equation are the basis elements of the source geometry.
The two additional basis elements of the ambient Clifford algebra are constructed by means of the first element of the paravector Clifford algebra $e_0=1$
\be
\label{basrest}
e_{n}=i \jmath\otimes 1
\,,\quad
e_{n+1}=\imath j\otimes 1\,.
\ee
The resulting basis elements generate the Clifford algebra $\mathbb{R}_{m+1,m+1}$. 
In contrast to \cite{Ulr17} the representation is given in terms of a tensor product. This is necessary to avoid ambiguities in the representation, which will become obvious in Section~\ref{spinmomop}.

\section{Start with the Clifford algebra $\mathbb{R}_{0,0}$}
\label{origin}
In \cite{Ulr17} the series of projective spaces introduced in the previous section started from the complex numbers, referring to the Clifford algebra $\mathbb{R}_{0,1}$.
However, from a conceptual point of view, it is more appealing to start from a null algebra consisting of the empty set.
This null algebra can be assigned to the Clifford algebra $\mathbb{R}_{0,0}$.
Nevertheless, one can construct a paravector algebra based on $\mathbb{R}_{0,0}$, which is made up of the trivial identity basis element $e_0=1$ alone.
The paravector algebra is thus representing the real numbers $\mathbb{R}$ as the elementary geometry, which can be identified with an originating point in a projective context.
The scheme introduced in the previous section can be applied to the paravector algebra $\mathbb{R}_{0,0}$.
There are no basis elements $e_k$ available in Eq.~(\ref{basspace}). However, two basis elements can be derived by means of Eq.~(\ref{basrest})
\be
e_1=i\jmath\otimes 1=\left(\begin{array}{cc}
0&i\\
i&0
\end{array}\right),\quad 
e_2=\imath j\otimes 1=\left(\begin{array}{cc}
0&j\\
-j&0
\end{array}\right).
\ee
Thus one arrives at a representation of the Clifford algebra $\mathbb{R}_{1,1}$.
The two basis elements can be used together with the identity to form a paravector model $e_\mu=(1,e_i)$ of the space $\mathbb{R}^{2,1}$.
The complex numbers, which form the initial algebra in \cite{Ulr17}, are included as the subalgebra $e_\mu=(1,e_1)$ representing the space $\mathbb{R}^{2,0}$.

\section{From the Clifford algebra $\mathbb{R}_{1,1}$ to $\mathbb{R}_{2,2}$}
With the Clifford algebra $\mathbb{R}_{1,1}$ in place, Eqs.~(\ref{basspace}) and (\ref{basrest}) can be applied again to generate the next higher dimensional geometry. 
One arrives now at the Clifford algebra $\mathbb{R}_{2,2}$ consisting of four matrices. 
Equation~(\ref{basspace}) results in the first two basis elements of $\mathbb{R}_{2,2}$
\be
\label{detspace}
e_1=\imath\jmath \otimes e_1\,,\quad 
e_2=\imath \jmath \otimes  e_2\,.
\ee
The elements $e_1$ and  $e_2$ on the left-hand side of the equations refer to the higher dimensional geometry. The elements $e_1$ and $e_2$  in the tensor product refer to the basis elements of the Clifford algebra $\mathbb{R}_{1,1}$.
Two further basis elements for $\mathbb{R}_{2,2}$ can be generated by means of Eq.~(\ref{basrest}). Now the identity element refers to a two-dimensional matrix
\be
\label{detrest}
e_3=i\jmath\otimes 1\,,\quad 
e_4=\imath j\otimes 1\,.
\ee
The four $4\times 4$ basis matrices correspond to a representation of the Clifford algebra $\mathbb{R}_{2,2}$ and can be considered as a paravector model for the geometric space $\mathbb{R}^{3,2}$.
The scheme given by Eqs.~(\ref{basspace}) and (\ref{basrest}) can then be applied again to generate $8\times 8$ matrices and so on.

\section{The spin tensor}
\label{spinmomop}
One can consider the algebra introduced in the previous section in more detail.
The product of two basis elements can be written in terms of symmetric and anti-symmetric contributions
\be
\label{basis}
e_\mu\bar{e}_\nu=g_{\mu\nu}+\sigma_{\mu\nu}\,.
\ee
Here the bar symbol $\bar{e}_\nu$ refers to conjugation of the considered basis element.
The symmetric contributions of the product are represented in terms of the metric tensor $g_{\mu\nu}$.
The anti-symmetric contributions are identified with the spin tensor $\sigma_{\mu\nu}$, see \cite{Ulr17} for more details.
The right hand side of Eq.~(\ref{basis}) does not expose any complex or hypercomplex units, which differs from the representation in previous publications of the author, e.g., in \cite{Ulr13}.

One can calculate now the spin tensor $\sigma_{\mu\nu}$ of the Clifford algebra $\mathbb{R}_{2,2}$. The result is displayed in Eq.~(\ref{confspin}).
The four elements in the first column of the matrix, which is given the index $0$, can be identified with the basis elements of the algebra $\sigma_{i0}=e_i$.
The other elements can be calculated by direct matrix multiplication.
Further insight is obtained if one breaks up the spin matrices into the basis elements of the base geometry by means of Eqs.~(\ref{detspace}) and (\ref{detrest}).
The calculation can be performed even more easily than with matrices.
One then obtains the spin tensor $\sigma_{\mu\nu}$ of the
Clifford algebra $\mathbb{R}_{2,2}$ in terms of the basis elements of the Clifford algebra $\mathbb{R}_{1,1}$ and the set of four hypercomplex units introduced in the previous sections
\be
\label{confspin}
\sigma_{\mu\nu}=\left(\begin{array}{ccccc}
0&-\imath\jmath \otimes e_1&-\imath\jmath \otimes e_2&-i\jmath\otimes 1&-\imath j \otimes 1\\
\imath\jmath \otimes e_1&0&1 \otimes \imath\jmath ij&-\imath i\otimes e_1&-\jmath j\otimes e_1\\
\imath\jmath \otimes e_2&-1 \otimes \imath\jmath ij&0&-\imath i\otimes e_2&-\jmath j\otimes e_2\\
i\jmath\otimes 1&\imath i\otimes e_1&\imath i\otimes e_2&0&\imath\jmath ij\otimes 1\\
\imath j\otimes 1&\jmath j\otimes e_1&\jmath j\otimes e_2&-\imath\jmath ij\otimes 1&0\\
\end{array}\right).
\ee
The result can be compared with the corresponding matrix for the Clifford algebra $\mathbb{R}_{2,1}$ in \cite{Ulr17}.
One finds that the third row $\sigma_{2\nu}$ as well as the third column $\sigma_{\mu 2}$ are inserted and the notation is enriched by the tensor product.
The tensor product is necessary, because the two diagonal matrices $\sigma_{12}=1 \otimes \imath\jmath ij$ and $\sigma_{34}=\imath\jmath ij\otimes 1$ can then be distinguished from each other.

\section{Commutation relations}
The spin angular momentum operator corresponds to the spin tensor divided by a factor of two
\be
\label{spin}
s_{\mu\nu}=\frac{\sigma_{\mu\nu}}{2}\,.
\ee
The commutation relations of the spin angular momentum operator can be taken from the literature.
They are summarized in the following sum of four terms
\be
\label{comm}
[s_{\mu\nu},s_{\rho\sigma}]=g_{\mu\sigma}s_{\nu\rho}-g_{\mu\rho}s_{\nu\sigma}-g_{\nu\sigma}s_{\mu\rho}
+g_{\nu\rho}s_{\mu\sigma}\,.
\ee
If one uses this formula and inserts the spin angular momentum operator given by Eqs.~(\ref{confspin}) and (\ref{spin}) the 
corresponding metric tensor can be calculated. One finds
\be
g_{\mu\nu}=\left(\begin{array}{ccccc}
1&0&0&0&0\\
0&1&0&0&0\\
0&0&-1&0&0\\
0&0&0&1&0\\
0&0&0&0&-1
\end{array}\right).
\ee
Thus the matrix generators are referring to the Anti-de Sitter group $SO(3,2,\mathbb{R})$.
Compared to \cite{Ulr17}, which referred at this stage to the Lorentz group $SO(3,1,\mathbb{R})$, the third column $g_{\mu 2}$ and the third row $g_{2\nu}$ have been added.

\section{Lie sphere geometry}
\label{Lie}
The space $\mathbb{R}^{3,2}$ can be represented as a paravector algebra with the Clifford algebra $\mathbb{R}_{2,2}$. It is this space, which is used to
represent Lie sphere geometry in two-dimensional manifolds like $\mathbb{R}^2$, the sphere $S^2$ or the hyperbolic space  $\mathbb{H}^2$. 
One finds that a line in the Lie quadric of $\mathbb{R}^{3,2}$ corresponds to a pencil of oriented spheres in $\mathbb{R}^2$. Coordinates in the space $\mathbb{R}^{3,2}$ thus represent these oriented spheres.
The representation includes also point spheres and planes as limit cases. For more details it is referred to Cecil \cite{Cec08} and Jensen et al. \cite{Jen16}.

In the considered five-dimensional space the metric consists of two timelike coordinates
assigned to $e_2$ and $e_4$. In order to match the metric with the Clifford algebra and identify the M\"obius geometry of the limit $r=0$ correctly, one has to shuffle the coordinates compared to Cecil \cite{Cec08}.
The two orientations of the sphere in $\mathbb{R}^2$ with center $p$ and unsigned radius $r>0$ are then represented by the two projective points
\be
\label{cecil}
\left[\left(p,\pm r,\frac{1-p\cdot p+r^2}{2},\frac{1+p\cdot p-r^2}{2}\right)\right]\,.
\ee
Here the notation has been chosen exactly as in \cite{Cec08} to allow for a straightforward comparison. 

Clifford paravectors can now be formed by contracting the basis elements of the Clifford algebra $\mathbb{R}_{2,2}$ with the above projective points.
The corresponding vector of basis elements can be written out in detail as
\be
e_\mu=(1,e_1,e_2,e_3,e_4)\,.
\ee
With this procedure the Lie sphere geometry is assigned to elements of the considered Clifford algebra.

\section{Group limit and momentum operators}
\label{plane}
The Poincar\'e group can be obtained from the Anti-de Sitter group $SO(3,2,\mathbb{R})$ as a group limit.
Following the discussion at the end of Section \ref{spinmomop} the third row and the third column is used to define the spin representation of the momentum operators in the limit of small $\epsilon$ as $p_\mu=\epsilon s_{\mu 2}$.
The spin angular momentum can be restricted to the remaining Minkowski subspace
\be
\label{grouplimit}
\lim_{\epsilon\rightarrow 0}\left(\begin{array}{ccccc}
0&s_{01}&\epsilon s_{02}&s_{03}&s_{04}\\
s_{10}&0&\epsilon s_{12}&s_{13}&s_{14}\\
\epsilon s_{20}&\epsilon s_{21}&0&\epsilon s_{23}&\epsilon s_{24}\\
s_{30}&s_{31}&\epsilon s_{32}&0&s_{34}\\
s_{49}&s_{41}&\epsilon s_{42}&s_{43}&0
\end{array}\right)
=\left(\begin{array}{ccccc}
0&s_{01}&p_{0}&s_{02}&s_{03}\\
s_{10}&0&p_{1}&s_{12}&s_{13}\\
-p_{0}&-p_{1}&0&-p_{2}&-p_{3}\\
s_{20}&s_{21}&p_{2}&0&s_{23}\\
s_{30}&s_{31}&p_{3}&s_{32}&0
\end{array}\right).
\ee
The considered groups have been applied also  in extensions of general relativity to the Anti-de Sitter group $SO(3,2,\mathbb{R})$,
see the comments and reprints in the corresponding section of Blagojevic and Hehl \cite{Bla13}.
The momentum $p$ should not be confused with the center of the Lie sphere in Eq.~(\ref{cecil}).

One should remember that now the Anti-de Sitter group is used to parametrize 
the Lie Sphere geometry in the two-dimensional Euclidean plane $\mathbb{R}^2$.
For example, the momenta in the ambient space correspond to the generators, which change the radius of the Lie spheres within the two-dimensional base geometry.
Whereas the momenta in the base geometry are a linear combination of the following angular momentum operators in ambient space
\be
p_0=-s_{03}-s_{04}\,,\quad
p_1=-s_{13}-s_{14}\,.
\ee
These relations have been used before for example by Kastrup \cite{Kas62} and they have been applied also in \cite{Ulr17}.
The Lie sphere geometry in the plane $\mathbb{R}^{2}$ may find applications for example in condensed matter physics \cite{Zaa15}.

\section{Extension to higher dimensional spaces}
With the scheme given by Eqs.~(\ref{basspace}) and (\ref{basrest}) one can extend to higher dimensional geometries.
One arrives at the Clifford algebra $\mathbb{R}_{3,3}$, which can be used as a paravector model for $\mathbb{R}^{4,3}$
\be
\label{finalbasis}
e_\mu=(1,e_1,e_2,e_3,e_4,e_5,e_6)\,.
\ee
The space $\mathbb{R}^{4,3}$ can be used to represent a Lie quadric referring to oriented hyperboloids, oriented point hyperboloids, and oriented hyperplanes in Minkowski space $\mathbb{R}^{3,1}$.
Thus, the center of the Lie sphere in Eq.~(\ref{cecil}) is an element of Minkowski space. The Lie spheres as they have been originally considered by Bateman and Timerding \cite{Bat10,Tim12} are thus a subgeometry.

\section{Application to nuclear physics}
\label{nuclear}
The basis elements of the Clifford algebra $\mathbb{R}_{3,3}$ are referring now to $8\times 8$ matrices. As these matrices are generated over the bicomplex numbers they are
in fact equivalent to $16\times 16$ complex matrices. 
The matrix structure can be used to incorporate Pauli spin $\times$ weak isospin $\times$ hadronic isospin $\times$ Dirac anti-particles states.
Each of the mentioned participants contributes with a factor of two to the overall spin structure. Thus, the model is eligible to consider for example the hadronic level including protons, neutrons, neutrinos, and electrons.
With the scheme given by Eqs.~(\ref{basspace}) and (\ref{basrest}) it is possible to invoke further higher dimensional geometries.
String theory even invokes a ten-dimensional space to have enough structure to cover the Standard Model 
with its three generations of quarks and leptons \cite{Gre87}.
However, in order to proof feasibility of the concept it seems to be better to take a step back and investigate nuclear processes
within effective quantum field theories on the hadron level \cite{Ser86}.

The motivation to work in spaces of a priori undetermined dimensionality is given by the hope 
that the method can identify the gauge symmetries in the sequence of geometric spaces to be of projective nature \cite{Dun10}, 
whereas global symmetries like the chiral symmetry are related to coordinate transformations of the geometrized internal symmetries in a space of fixed dimension.
Note that the weak isospin is related to a local gauge symmetry, whereas the hadronic isospin is described by a global symmetry.
In this context one could try to relate the corresponding symmetry breaking mechanisms, which lead to either Higgs or Goldstone bosons, with a geometric perspective
in order to derive particle masses. This thought can be motived by a hypothetical mass formula for electrons and protons \cite{Ulr17}
\be
\label{proton}
4\pi\exp{(4\pi)}=\left(\frac{m_p}{m_e}\right)^2\,.
\ee
One may assume that the left-hand side of the equation is related to an angular momentum. The right-hand side is given in terms of a squared mass ratio.
The formula is thus inspired by the relation between angular momentum and squared masses given by the Regge trajectories.
The experimental proton to electron mass ratio is calculated from this mass formula with a deviation of $3.4\%$.
If one understands the $4\pi$ being related to the surface area of a unit sphere, a minor deformation towards an ellipsoid can change the surface area in a way that the values 
on both sides of Eq.~(\ref{proton}) match exactly.

\section{Scattering amplitudes in a Klein-Gordon theory for fermions}
\label{scatt}
Beside such fundamental considerations one can investigate the question, whether it is possible to reparametrize existing experimental data
with the algebraic expressions introduced in the previous sections. The intention is to work with a Klein-Gordon theory for fermions.
For spinless particles the Klein-Gordon theory provides scattering amplitudes of the form \cite{Bjo65,Nac86}
\be
\label{neuelec}
\langle f \vert \, J_\mu\,\vert i \rangle =p_{i\mu}+p_{f\mu}\;.
\ee
Here $p_i$ is a Minkowski space vector, which describes the momentum of the initial state and $p_f$ the momentum of the outgoing state.
In a fermion theory based on the Klein-Gordon equation
the matrix elements of Eq.~(\ref{neuelec}) will be replaced by a current of the form
\be
\label{scatmat}
\langle f \vert \, J_\mu\,\vert i \rangle =\bar{u}_f(p_i^\nu e_\nu\bar{e}_\mu+e_\mu\bar{e}_\nu p_f^\nu)u_i\;.
\ee
This expression has been applied in \cite{Ulr13} to Mott scattering and it has been shown that the scattering amplitude is equivalent to the corresponding amplitude obtained in the Dirac theory.

The correspondence to the Dirac theory can be formally confirmed quite easily.
Equation (\ref{scatmat}) can be transformed with the help of Eq.~(\ref{basis}) into a representation, which is proportional to the Gordon decomposition of the Dirac theory
\be
\label{tracescamp}
\langle f \vert \, J_\mu\,\vert i \rangle =\bar{u}_f(d_\mu+\sigma_{\mu\nu}q^\nu) u_i \,.
\ee
The initial and final momenta are encoded in the following expressions
\be
\label{momenta}
d=p_f+p_i\;,\hspace{1cm}q=p_f-p_i\;.
\ee
One can now generalize these expressions to higher dimensional geometries.
The initial and final momenta are now supposed to live in the momentum space corresponding to the Lie sphere geometry in $\mathbb{R}^{4,3}$.
The $u_i$ and $u_f$ are then eight-component spinors.
As the spin tensor $\sigma_{\mu\nu}$ is evaluated with the basis elements of the Clifford algebra $\mathbb{R}_{3,3}$ given by Eq.~(\ref{finalbasis}), the matrix elements of the spin tensor are bicomplex $8\times 8$ matrices.
In analogy to Eq.~(\ref{confspin}) it is again possible to break up $\sigma_{\mu\nu}$ to further evaluate its substructure. Now the subcomponents are referring to the basis elements of the Clifford algebra $\mathbb{R}_{2,2}$.

One can now make first attempts to reparametrize existing experimental data in terms of the spin group of $SO(4,3,\mathbb{R})$ and interpret the results within Minkowski space $\mathbb{R}^{3,1}$.
Consider again the discussion in Section \ref{plane} within the corresponding lower dimensional geometry, where $SO(3,2,\mathbb{R})$ is replaced now by $SO(4,3,\mathbb{R})$
and $\mathbb{R}^2$ by $\mathbb{R}^{3,1}$.

\section{Summary}
The modulo $(1,1)$ periodicity of Clifford algebras has been applied starting from the initial
Clifford  algebra $\mathbb{R}_{0,0}$. The sequence of higher dimensional Clifford algebras $\mathbb{R}_{n,n}$, generated by inverse projections, is
related to Lie sphere geometries, which naturally include M\"obius geometries as their subgeometries.
An essential part in this representation play the bicomplex numbers, which stand in relation to $S^3$ in the same way 
as the complex numbers stand in relation to $S^1$. The diffeomorphism of the bicomplex numbers to $SU(2,\mathbb{C})$ and the relation to spinors is briefly discussed.

It is indicated how the higher dimensional Clifford algebras can be assigned on a qualitative level to the internal particle symmetries.
In an initial model the hadronic isospin can be combined with the weak isospin, the electromagnetic spin,
and the Dirac negative energy contributions to form a geometrized model of the overall spin and isospin structure.
The considered Clifford algebra is $\mathbb{R}_{3,3}$. As the sequence of Clifford algebras is unlimited further
higher dimensional geometries can be invoked with the given scheme to include further internal particle symmetries.

\end{document}